\documentclass[sigconf]{acmart}

\usepackage{multirow}
\usepackage{graphicx}

\AtBeginDocument{%
  \providecommand\BibTeX{{%
    \normalfont B\kern-0.5em{\scshape i\kern-0.25em b}\kern-0.8em\TeX}}}

\copyrightyear{2022} 
\acmYear{2022} 
\setcopyright{rightsretained} 
\acmConference[WWW '22 Companion]{Companion Proceedings of the Web Conference 2022}{April 25--29, 2022}{Virtual Event, Lyon, France}
\acmBooktitle{Companion Proceedings of the Web Conference 2022 (WWW '22 Companion), April 25--29, 2022, Virtual Event, Lyon, France}\acmDOI{10.1145/3487553.3524249}
\acmISBN{978-1-4503-9130-6/22/04}

\begin{document}

%%
%% The "title" command has an optional parameter,
%% allowing the author to define a "short title" to be used in page headers.
\title{PREP: Pre-training with Temporal Elapse Inference for Popularity Prediction}

%%
%% The "author" command and its associated commands are used to define
%% the authors and their affiliations.
%% Of note is the shared affiliation of the first two authors, and the
%% "authornote" and "authornotemark" commands
%% used to denote shared contribution to the research.
\author{Qi Cao$^{1}$, Huawei Shen$^{1,3*}$, Yuanhao Liu$^{1,3}$, Jinhua Gao$^{1}$, Xueqi Cheng$^{2,3}$}
\affiliation{%
  \institution{\{caoqi, shenhuawei, liuyuanhao20z, gaojinhua, cxq\}@ict.ac.cn}
  \institution{$^1$ Data Intelligence System Research Center, Institute of Computing Technology, Chinese Academy of Sciences, China}
  \institution{$^2$ CAS Key Laboratory of Network Data Science and Technology, Institute of Computing Technology, Chinese Academy of Sciences, China}
  \country{$^3$ University of Chinese Academy of Sciences, China}
}

%%
%% By default, the full list of authors will be used in the page
%% headers. Often, this list is too long, and will overlap
%% other information printed in the page headers. This command allows
%% the author to define a more concise list
%% of authors' names for this purpose.
\renewcommand{\shortauthors}{Qi Cao, et al.}

%%
%% The abstract is a short summary of the work to be presented in the
%% article.
\begin{abstract}
Predicting the popularity of online content is a fundamental problem in various applications. One practical challenge takes roots in the varying length of observation time or prediction horizon, i.e., a good model for popularity prediction is desired to handle various prediction settings. 
However, most existing methods adopt a separate training paradigm for each prediction setting and the obtained model for one setting is difficult to be generalized to others, causing a great waste of computational resources and a large demand for downstream labels. To solve the above issues, we propose a novel \emph{\textbf{pre}-training framework for \textbf{p}opularity prediction}, namely \emph{\textbf{PREP}}, aiming to pre-train a general representation model from the readily available unlabeled diffusion data, which can be effectively transferred into various prediction settings. We design a novel pretext task for pre-training, i.e., \emph{temporal elapse inference} for two randomly sampled time slices of popularity dynamics, impelling the representation model to learn intrinsic knowledge about popularity dynamics. Experimental results conducted on two real datasets demonstrate the generalization and efficiency of the pre-training framework for different popularity prediction task settings.
\let\thefootnote\relax\footnotetext{*Corresponding Author}
\end{abstract}

%%
%% The code below is generated by the tool at http://dl.acm.org/ccs.cfm.
%% Please copy and paste the code instead of the example below.
%%


%%
%% Keywords. The author(s) should pick words that accurately describe
%% the work being presented. Separate the keywords with commas.

%%
%% This command processes the author and affiliation and title
%% information and builds the first part of the formatted document.
\maketitle

\section{Introduction}

\begin{figure}
  \centering
  \includegraphics[width=2.7in]{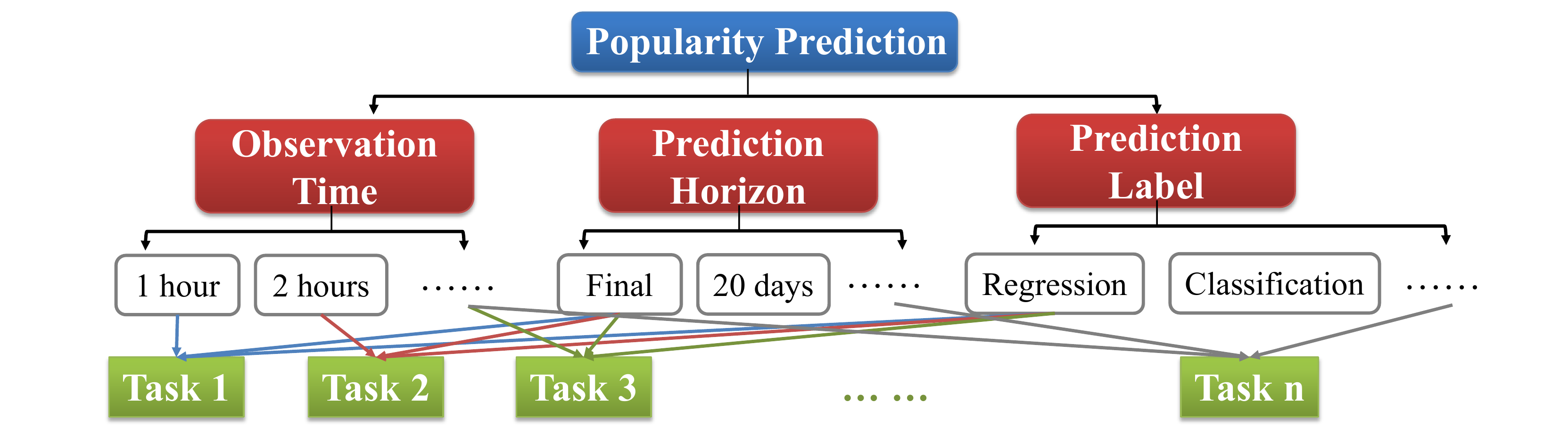}
  
\hspace{0.4cm}(a) Various Popularity Prediction Settings

  \includegraphics[width=2.8in]{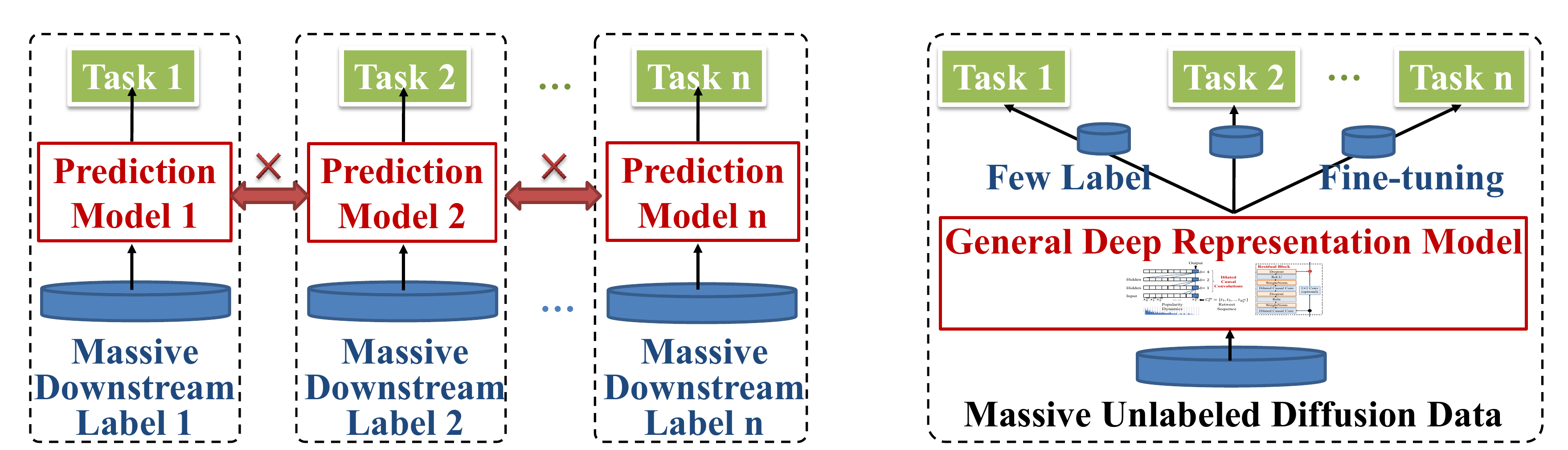}
  
\hspace{0.5cm}(b) Separate Training  \hspace{1.0cm}  (c) Pre-training Framework
    \caption{Popularity prediction tasks and comparison between \emph{Separate Training} and \emph{Pre-trianing Framework}.}
  \Description{motivation}
  \label{fig:motivation}
\end{figure}

The prevalence of social platforms, e.g., Twitter, Sina Weibo, has brought great convenience for the production and dissemination of user-generated online content. Every day there are tens of millions of online content generated on these platforms~\cite{Zhao2015Seismic,Cao2017Deephawkes}. Faced with such a large amount of information, predicting the future popularity of online content in advance plays an important role in various applications~\cite{Tatar2014Survey,Gao2019Taxonomy}, e.g., social recommendation, online advertisement, information retrieval.

One practical challenge for popularity prediction takes roots in the different settings of popularity prediction tasks in different situation~\cite{Hofman486}, shown in Figure~\ref{fig:motivation} (a). 
Specifically, there may be different settings of the observation time window, varying from 1 hour (predict the future popularity with an observation time of 1 hour) to 2 hours or other more~\cite{Shao2019TCN,Chen2019CasCN,Cao2020CoupledGNN,SIGIR21}, while there may also be different prediction horizons~\cite{Zhao2015Seismic,Cao2017Deephawkes,Gao2019Taxonomy}. Even the type of prediction label may change from binary classification (e.g., predict whether the popularity will double in the future)~\cite{Cheng2014Feature,Liao2019DTCF} to regression (predict exact future popularity)~\cite{Li2017Deepcas,Cao2017Deephawkes,Chen2019CasCN}. Such a situation brings great challenges to practical application, i.e., a good popularity prediction model is desired to handle various prediction task settings.

Existing methods for popularity prediction mainly fall into three categories~\cite{Gao2019Taxonomy}: feature-based methods, generative methods, and deep learning based methods. 
Feature-based methods generally extract various hand-crafted features for popularity prediction~\cite{Tatar2014Survey,Cheng2014Feature}, while generative methods regard the popularity dynamics as an arrival point process and model the intensity function by different assumptions~\cite{Zhao2015Seismic, Mishra2016FeatureDriven}. The performance of these methods heavily depends on the heuristically extracted features or the unknown assumption, limiting their prediction performance.
Recently, deep learning based methods have emerged and achieved state-of-the-art prediction performance~\cite{Li2017Deepcas, Cao2017Deephawkes, Du2016RMTPP, deepinf2018, Cao2020CoupledGNN, Shao2019TCN, SIGIR21}, which train a separate model for each prediction task under the guidance of downstream labels (Figure~\ref{fig:motivation} (b)).
The obtained model for one prediction task setting is difficult to be generalized to other task settings, causing a great waste of training time and computational resources, as well as a large demand for downstream labels. 

To inherit the powerful ability of deep learning based methods while eliminating the limitation of separate training paradigm, we propose a novel \emph{pre-training framework for popularity prediction}, see Figure~\ref{fig:motivation} (c). Instead of training a separate prediction model through massive downstream labels for each task setting, the proposed framework aims to pre-train a general representation model from readily available unlabeled diffusion data, which can be effectively transferred into different popularity prediction tasks.
As the key of pre-training framework mainly lies in the design of self-supervised pretext task, we propose a novel pretext task for pre-training, i.e., \emph{temporal elapse inference for two randomly sampled time slices of popularity dynamics}. Such a designed pretext task enforces the deep model to capture the intrinsic evolution pattern of popularity dynamics, so as to benefit various downstream task settings. 

Note that, the pre-trained representation model only needs to be fine-tuned by few downstream labels when transferred into different downstream settings. Experiments conducted on both Sina Weibo and Twitter demonstrate that when compared with the prediction model under a separate training paradigm, the proposed framework is much more efficient and generalizable while achieving comparable performance. When compared with the random initialization, the pre-trained representation model achieves significant improvement on downstream popularity prediction tasks, further demonstrating the effectiveness of the pre-training framework.

\section{Methods}
Since temporal information is the dominant factor for popularity prediction~\cite{Cheng2014Feature} and can be easily generalized across different platforms, here we focus on \emph{time-aware popularity prediction} scenario.

\emph{\textbf{Time-aware popularity prediction task $\mathcal{T}$:} Given the observed retweet sequence of online content $m$ within observation time $T$, i.e., $C^m_T = \{t_1,t_2,...,t_{N^m_T}\}$ where $N^m_T$ is the total number of retweets, it aims to predict the popularity label $y_m$ at a prediction horizon $T_p$.}

Different settings of observation time $T$, prediction horizon $T_p$, and popularity label $y_m$ form different popularity prediction tasks.

\subsection{Overview of Pre-training Framework}
\label{sec:downstreamtasks}
Given a set of popularity prediction tasks $\{\mathcal{T}_i\}_{i=1}^{N_{tasks}}$ and a set of corresponding datasets with massive task labels $\{\mathcal{D}^{mas}_i\}_{i=1}^{N_{tasks}}$, existing paradigm trains a separate prediction model $f_{\theta_i}$ using $\mathcal{D}^{mas}_{i}$ for each task $\mathcal{T}_i$, which is both computational resources-consuming and massive label-demanded. 

In contrast, the \emph{pre-training for popularity prediction} aims to pre-train one general deep representation model $f_{\theta}$ using unlabeled diffusion data $\mathcal{D}$ based on a pretext task $\mathcal{T}_{pre}$, such that the pre-trained model $f_{\theta}$ can be effectively transformed into various (unseen) downstream popularity prediction tasks $\{\mathcal{T}_i\}_{i=1}^{N_{tasks}}$, via fine-tuned by few downstream labels $\mathcal{D}^{few}_{i}$.
In this paper, we take the superior Temporal Convolutional Neural Networks (TCN)~\cite{Bai2018Empirical,Shao2019TCN} as the base deep model and transfer the input retweet sequence $C^m_T$ into popularity dynamics $X^m_T=[x^m_1,x^m_2,...,x^m_T]$, which describes the incremental popularity $x^m_i$ per time unit $i$ to serve as the input of TCN. 

\subsection{Pretext Taks: Temporal Elapse Inference}
\begin{figure}
  \centering
  \includegraphics[width=2.65in]{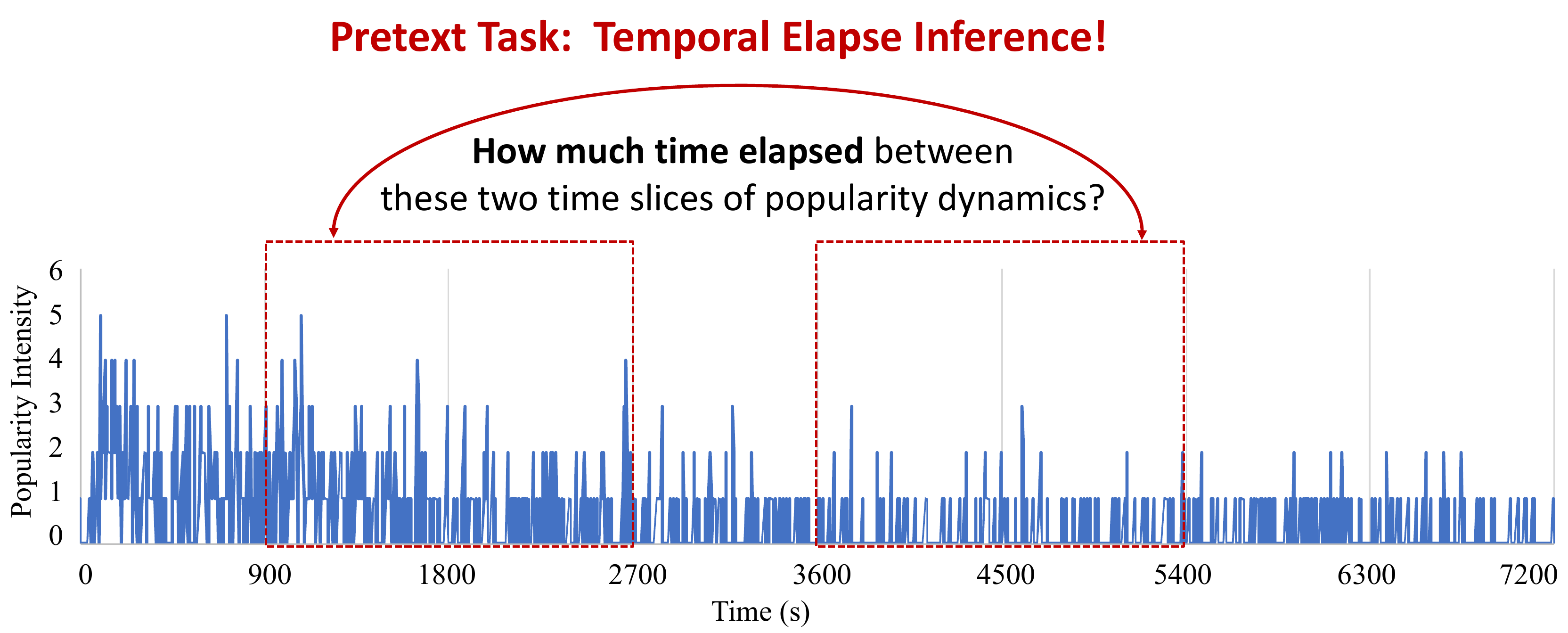}
  \caption{Illustration of temporal elapse inference. The curve depicts the incremental popularity per time unit on a real example popularity dynamics in Sina Weibo.}
  \Description{framework}
  \label{fig:framework}
\end{figure}

To learn a satisfactory general representation model, the key lies in the design of the pretext task $\mathcal{T}_{pre}$. Considering that the popularity dynamics may have fluctuations in each time slice but remains relatively stable in temporal evolution, we propose a novel \emph{temporal elapse inference (TEI)} as the pretext task. TEI randomly samples pairs of time slices of popularity dynamics and aims to infer the time elapsed between these two time slices, see Figure~\ref{fig:framework} to have an intuitive understanding. 
In order to accurately predict the temporal elapse between two time slices, the deep representation model needs to understand the temporal context information and capture the evolution pattern of popularity dynamics varying with time. Such ability is critical for downstream popularity prediction tasks, which is the reason why the pre-trained deep representation model can be beneficial to downstream tasks. Next, we formally define the designed pretext task of temporal elapse inference.

\subsubsection{Temporal Context Sampling}
\label{sec:templecontextsampling}
We first segment the input popularity dynamics into several time slices. Let $\Delta T$ denote the length of each time slice, then the popularity dynamics of online content $m$ can be segmented as:
$\{X^{m}_{T,1} = [x_1,...,x_{\Delta T}]$, ... , $X^m_{T,s} = [x_{(s-1)\cdot \Delta T+1},$ $...,x_{s\cdot \Delta T}]\}$,
where $T$ is the length of observation time, $s = \lfloor T/\Delta T \rfloor$ denotes the total number of time slices, and $X^m_{T,i}$ denote the $i$-th time slice of popularity dynamics. Let $l_{e}$ denote the temporal elapse between the $A-$th time slice and $B-$th time slice, i.e., $l_e = B-A$. 

Intuitively, two time slices that are too far away may make the temporal elapse inference too difficult to confuse the deep representation model, while two overlapping time slices may result in a simple prediction problem that can be easily solved without learning any general knowledge. Based on the above intuition, we set a maximum temporal elapse $l_{max}$, and then uniformly sample the temporal elapse $l_e\sim P_l, \text{where } p_l(i) = \frac{1}{min(s,l_{max})},i=1,2,...,min (s,l_{max})$.
To make sure that the sampled time slices contain sufficient observation, we assign higher sampling probability to earlier time slices, i.e., the $A$-th time slice is sampled with the probability $p_a(A) \propto f(A)$, where $f(\cdot)$ is monotone decreasing, and then $B=l_e+A$.

\subsubsection{Temporal Elapse Inference}
For the sampled pair of time slice $A$ and $B$, we apply TCN with $L$ layers~\cite{Bai2018Empirical,Shao2019TCN} on the two time slices of popularity dynamics $X^m_{T,A}$ and $X^m_{T,B}$ respectively to obtain their representations, i.e., 

\begin{equation}
o^{m,j} =  TCNConv_L\left(...\left(TCNConv_1\left(X^m_{T    ,j}\right)\right)...\right), j=A,B.
\end{equation}

Then the temporal elapse is predicted as:
\begin{math}
\hat{l}_e = MLP_{p} \left( o^{m,A}||o^{m,B}\right),
\end{math}
where $||$ denotes the operator of vector concat, and $MLP_{p}$ means a multi-layer perceptron.
The entire model is pre-trained by receiving signals from the real temporal elapse $l_e$, i.e.,
\begin{equation}
Loss = (l_e-\hat{l}_e)^2.
\end{equation}

\subsection{Transfer into Downstream Tasks}
\paragraph{Downstream Tasks} For downstream popularity prediction tasks defined in Section~\ref{sec:downstreamtasks}, we transfer the pre-trained model on the observed popularity dynamics within time $T$, i.e., $X^m_T=[x^m_1,x^m_2,...,x^m_T]$. Formally,
\begin{math}
o^m = TCNConv_L(...(TCNConv_1(X^m_T))...)
\end{math}
and the predicted label is:
\begin{math}
\hat{y}_m = MLP_{d}\left( o^m\right ), 
\end{math}
where $MLP_{d}$ is a multi-layer perceptron.

\paragraph{Freezing vs Full fine-tuning}We offer two fine-tuning strategies for downstream popularity prediction tasks, i.e., freezing mode and full fine-tuning mode. The former freezes the parameters of the pre-trained TCN and treats it as a static representation extractor while only updating the parameters of $MLP_{d}$ on each downstream task, referred to as \textbf{PREP-TCN-f}. The latter mode updates all parameters on downstream tasks, referred to as \textbf{PREP-TCN}.

\section{Experiments}
We conduct experiments on two real datasets for various task settings. The code is publicly available in Github~\footnote{https://github.com/CaoQi92/PREP}.

\subsection{Experimental Setup}

\subsubsection{Datasets}
We experiment with two real datasets. The first is \textbf{Sina Weibo}, where we collect all the original messages produced between June 1, 2016 and June 10, 2016, containing 710,554 online content in total. 
The second is \textbf{Twitter}~\cite{Zhao2015Seismic}, containing 166,076 tweets in total. We sort all the online content by their publication time, and take the first 75\% for training, 15\% for validation, and the last 10\% as test set following~\cite{Cao2017Deephawkes,Zhao2015Seismic}.

\subsubsection{Downstream Popularity Prediction Tasks}
We vary the observation time, prediction horizon as well as the popularity label, forming four representative tasks $\mathcal{T}_1$, $\mathcal{T}_2$, $\mathcal{T}_3$, $\mathcal{T}_4$, see Table~\ref{tab:tasks}. 
For training loss, we take mean relative square error~\cite{Tatar2014Survey,Cao2020CoupledGNN}: 
\begin{math}
\text{MRSE} = \frac{1}{M}\sum_{m=1}^M\left( \frac{y_m-\hat{y}_m}{y_m}\right)^2
\end{math}
for regression tasks, and take the binary cross entropy for classification task.

\subsubsection{Baselines}
We choose state-of-the-art methods for time-aware popularity prediction as strong baselines, i.e., \textbf{Feature-based}~\cite{Cheng2014Feature}; \textbf{Seismic}~\cite{Zhao2015Seismic} as a typical generative method; \textbf{DeepHawkes}~\cite{Cao2017Deephawkes}, \textbf{CasCN}~\cite{Chen2019CasCN},  \textbf{TCN}~\cite{Shao2019TCN} as powerful deep learning based methods.

\subsubsection{Evaluation Metrics}
We adopt two commonly used evaluation metrics for regression task, i.e., loss function \textbf{MRSE}~\cite{Tatar2014Survey,Cao2020CoupledGNN}, and \textbf{R-Acc}~\cite{Gao2019Taxonomy} that measures the fraction of content that are correctly predicted under a given tolerance of error: $\frac{1}{M}\sum_{m=1}^M \mathbb{I}\left[ \text{APE}_m\le \epsilon\right]$ where $\text{APE}_m=\left| \frac{y_m-\hat{y}_m}{y_m}\right|$ and $\epsilon = 0.3$. As for classification task, we take the widely used evaluation metrics for binary classification, i.e., classification accuracy (denoted as \textbf{C-Acc}) and \textbf{F1} score.

\subsubsection{Implementation Details}
The hyper-parameters are tuned to obtain the best results on validation and choose learning rate from $\{10^{-5},5\times 10^{-5},10^{-4}..., 10^{-2}\}$, maximum temporal elapse $l_{max}$ from $\{6,12,18,24\}$. The length of each time slice $\Delta T$ equals 1800 seconds. For the base TCN model, kernel size $K=8$, number of layers $L=12$, time unit equals $5$ seconds, and the number of hidden units equals $8$. We use a mini-batch of 32 and stop training as long as the loss of validation doesn't decline for 50 consecutive iterations.

\begin{table}
  \caption{Downstream Popularity Prediction Tasks}
  \label{tab:tasks}
  \scalebox{0.7}{
  \begin{tabular}{ccccl}
    \toprule
    Tasks  & Observation Time$^1$ & Prediction Horizon$^1$ & Label$^2$\\
    \midrule
    $\mathcal{T}_1$    & 1 hour               &3 days / 1 days  &$\mathcal{R}$ \\
    $\mathcal{T}_2$    & 1 hour               &final &$\mathcal{R}$\\ 
    $\mathcal{T}_3$    & 2 hours              &final &$\mathcal{R}$\\
    $\mathcal{T}_4$    & 2 hours /0.5 hour    &final &$\mathcal{C}$\\

  \bottomrule
\end{tabular}}

\footnotesize{$^1$ The left of "/" means time for Weibo, while the right of "/" means time for Twitter}\\
\footnotesize{$^2$ \textbf{$\mathcal{R}$}: Regression task, predict the popularity within prediction horizon; \quad\quad\quad\quad\quad\quad}\\
\footnotesize{\textbf{$\mathcal{C}$}: Classification task, predict whether popularity double at the prediction horizon}
\end{table}

\subsection{Effectiveness of Pre-training Framework}
We conduct experiments with various downstream popularity prediction tasks with the following observations, see Table~\ref{tab:overallprediction}:

$\bullet$ \textbf{For separate training paradigm with massive labels, TCN shows outstanding performance on all downstream popularity prediction task settings,} which is consistent with the reported results in~\cite{Shao2019TCN}. 
We cannot perform DeepHawkes and CasCN on Twitter since this dataset lacks the structure information of diffusion subgraph.
Since Seismic is sensitive to outliers and can only predict the final popularity, we omit results of MRSE and task $\mathcal{T}_1$.

$\bullet$ \textbf{When transferring the pre-trained model into downstream tasks with few labels}, i.e., 0.1\% downstream labels in Sina Weibo and 0.5\% in Twitter, \textbf{our pre-trained TCN model significantly outperform the random initialized TCN model.} That is, PREP-TCN-f significantly outperforms TCN-f, and PREP-TCN also significantly outperforms TCN, demonstrating the effectiveness of our pre-training framework for downstream tasks.

$\bullet$ \textbf{The PREP-TCN which is fine-tuned with few downstream labels even achieves comparable prediction performance when compared with TCN trained with massive downstream labels under the paradigm of separate training.} For example, $0.232$ vs $0.238$ MRSE and $47.4\%$ vs $47.8\%$ R-Acc for task $\mathcal{T}_1$ on Sina Weibo. 
However, the separate training of TCN for various downstream prediction settings is much more time resource-consuming than the pre-training framework (See section~\ref{sec:trainingtime}).

\begin{table}
  \caption{Performance on Popularity Prediction Tasks}
  \label{tab:overallprediction}
  \scalebox{0.75}{
  \begin{tabular}{c|c|c|c|c|c|c|c|c}
    \toprule
     \multicolumn{1}{c| }{} & \multicolumn{2}{c| }{Task $\mathcal{T}_1$ }& \multicolumn{2}{c| }{Task $\mathcal{T}_2$} & \multicolumn{2}{c| }{Task $\mathcal{T}_3$  }& \multicolumn{2}{c }{Task $\mathcal{T}_4$}\\
    \midrule
      \multicolumn{1}{c| }{Methods} & MRSE & R-Acc 
      & MRSE &R-Acc
      & MRSE &R-Acc 
      & C-Acc  & F1  \\
    \midrule
    \midrule
     \multicolumn{9}{c }{Separate Training on Weibo with Massive Labels}\\
     \midrule
            Seismic 
     &-       &-
     &-       &35.1\%
     &-      &37.5\%
     &52.4\%  &0.508\\
    DeepHawkes 
    &0.510  &35.7\%
    &0.379  &38.9\%
    &0.342  &40.7\%
    &49.8\%  &0.000\\
    CasCN
    &0.347  &40.8\%
    &0.372  &38.4\%
    &0.326  &44.5\%
    &65.4\%  &0.664\\
    Feature-based 
    &0.251  &42.6\%
    &0.212  &43.9\%
    &0.172  &53.6\%
    &59.2\%  &0.690\\
               TCN 
    &\bf{0.232}  &\bf{47.4\%}
    &\bf{0.175}  &\bf{52.4\%}
    &\bf{0.137}  &\bf{63.1\%}
    &\bf{73.6\%}  &\bf{0.713}\\
    \midrule
         \multicolumn{9}{c }{Transfer Pre-trained (or random initialized) model on Weibo with Few Labels}\\
     \midrule
    TCN-f 
    &0.809  &0.4\%
    &0.396  &25.2\%
    &0.396  &25.2\%
    &49.7\%  &0.000\\
                  PREP-TCN-f  
    &0.322  &33.5\%
    &0.258  &40.1\%
    &0.236  &44.1\%
    &66.6\%  &0.645\\
                                TCN
    &0.262  &43.8\%
    &0.191  &51.0\%
    &0.168  &57.0\%
    &68.1\%  &\bf{0.674}\\
                               PREP-TCN 
    &\bf{0.238}  &\bf{47.8\%}
    &\bf{0.184}  &\bf{51.7\%}
    &\bf{0.147}  &\bf{61.4\%}
    &\bf{70.9}\%  &0.669\\ 
    
    \midrule
    \midrule
         \multicolumn{9}{c }{Separate Training on Twitter with Massive Labels}\\
     \midrule
               Seismic
    &-       &-
    &-       &60.7\%
    &-       &66.4\%
    &61.4\%  &0.520\\
     Feature-based
    &0.077  &77.8\%
    &0.106  &70.6\%
    &0.084  &77.9\%
    &65.3\%  &0.582\\
               TCN
    &\bf{0.054}  &\bf{82.3\%}
    &\bf{0.086}  &\bf{74.3\%}
    &\bf{0.063}  &\bf{81.9\%}
    &\bf{70.9\%}  &\bf{0.634}\\
    
    \midrule
             \multicolumn{9}{c }{Transfer Pre-trained (or random initialized) model on Twitter with Few Labels}\\
     \midrule
    TCN-f 
    &0.238  &40.7\%
    &0.258  &37.7\%
    &0.258  &37.7\%
    &54.8\%  &0.000\\
                                PREP-TCN-f
    &0.166  &53.1\%
    &0.192  &48.0\%
    &0.217  &46.1\%
    &65.6\%  &0.534\\
                                TCN
    &0.073  &76.1\%
    &0.100  &70.6\%
    &0.084  &76.7\%
    &\bf{70.7\%}  &0.614\\
                               PREP-TCN
    &\bf{0.057}  &\bf{83.0\%} 
    &\bf{0.090}  &\bf{71.9\%}
    &\bf{0.069}  &\bf{79.9\%}
    &70.6\%  &\bf{0.630}\\
    \bottomrule
  \end{tabular}}
\end{table}

\subsection{Efficiency of Pre-training Framework}
\label{sec:trainingtime}
\begin{figure}
  \centering
  \includegraphics[width=3.3in]{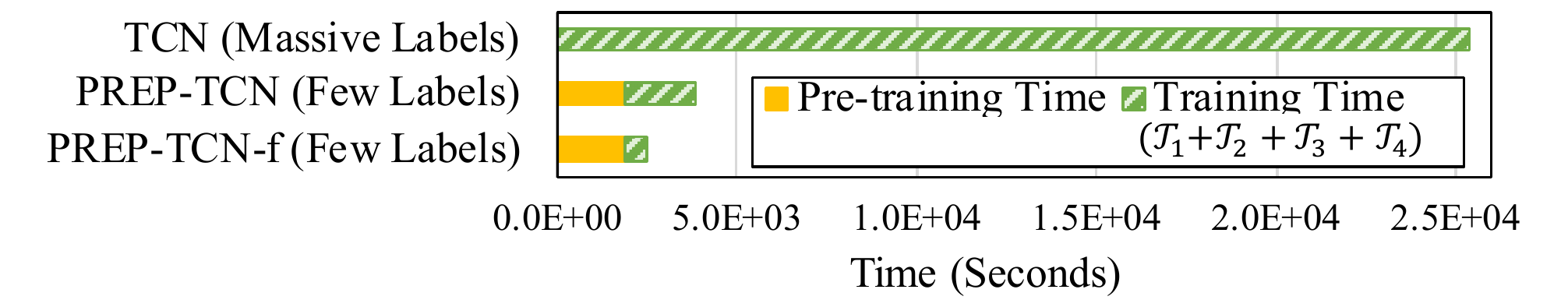}
  \caption{Efficiency Comparison}
  \Description{TimeAnalysis}
   \label{fig:TimeAnalysis}
\end{figure}

We conduct time experiments on a single GPU (NVIDIA Tesla K80) and first analyze the training time on downstream tasks, shown in Figure ~\ref{fig:TimeAnalysis}. Even taking into account the time of pre-training, PREP-TCN and PREP-TCN-f is much more efficient than the separately trained TCN with massive downstream labels. Such efficiency advantages of the pre-training framework will be more significant with the increase of the number of tasks. 

To deeper the understanding of efficiency advantages of the pre-training framework, we further analyze whether it can accelerate the convergence of model training.
Figure~\ref{fig:convergence} (left) shows that the loss of the pre-trained model decreases quickly at early training steps and then gradually tends to be stable, while the random initialized TCN convergence very slowly, demonstrating the benefits of the pre-training framework for model convergence.

\subsection{Analysis of Pre-training Task}
To demonstrate the superiority of temporal elapse inference (TEI) as the pre-training task, we conduct experiments with replaced pre-training tasks, i.e., take one of the downstream tasks $\mathcal{T}_1$ as the pre-training task. Besides, we also replace the sampling strategies in TEI with purely random sampling. 

\begin{figure}
  \centering
  \includegraphics[width=1.65in]{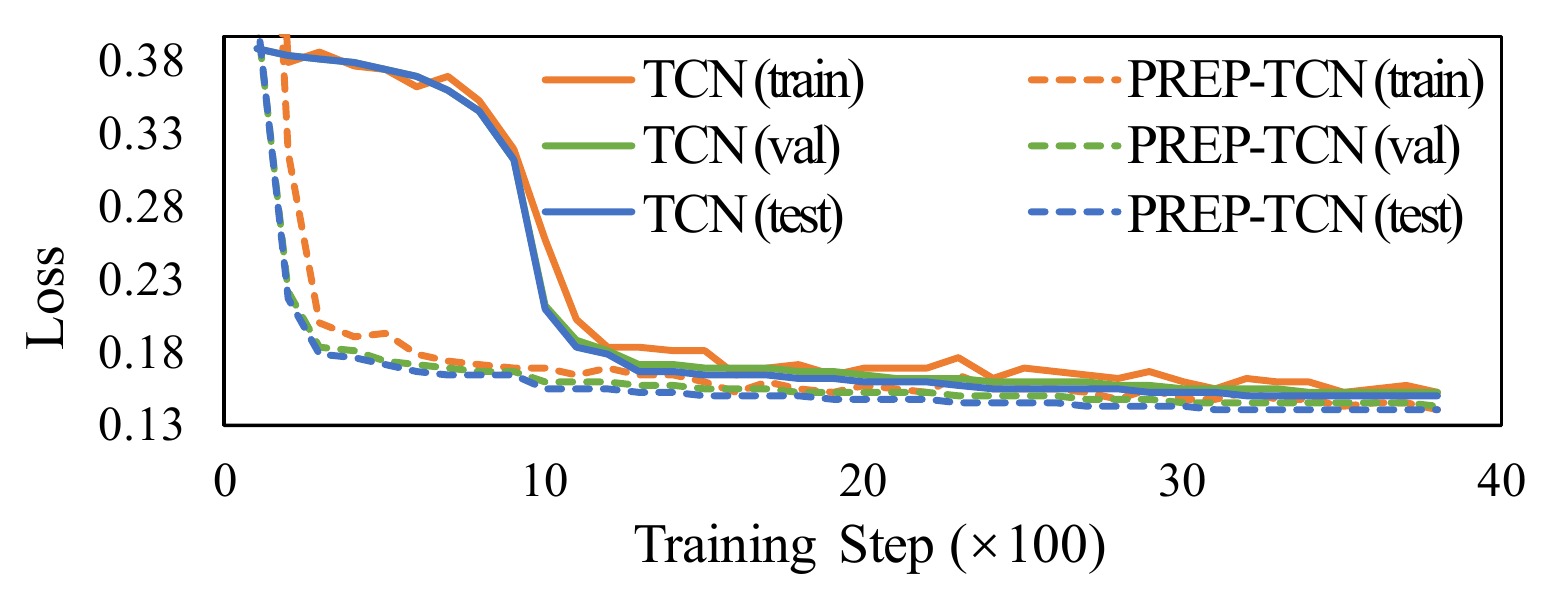}
    \includegraphics[width=1.65in]{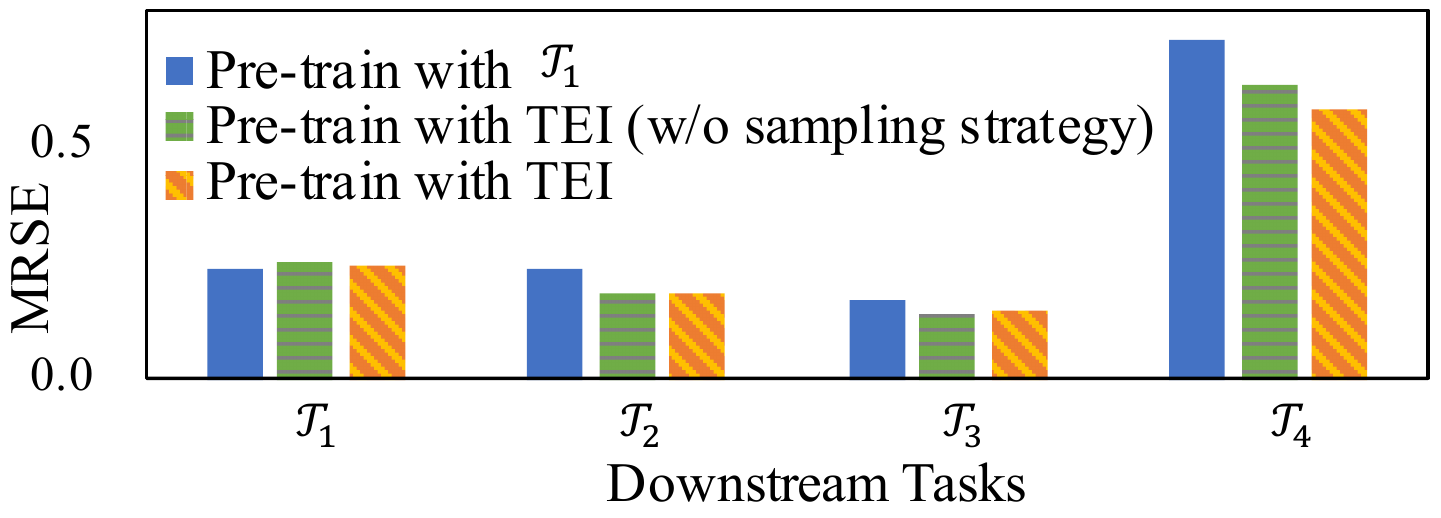}
  \caption{Left: Convergence Comparison; Right: Analysis of Pretext Task}
  \Description{convergence}
   \label{fig:convergence}
\end{figure}

Experimental results are shown in Figure~\ref{fig:convergence} (right). For model pre-trained with task $\mathcal{T}_1$, it achieves a comparable prediction performance when the downstream tasks exactly match the pre-training task, i.e., the downstream task is also $\mathcal{T}_1$, but achieves poor prediction performance for other downstream tasks. These results show that it is difficult to transfer or generalize the learned model across different downstream tasks. When removing the designed sampling strategy, the pre-trained model performs not as well as the original TEI, validating the effectiveness of the sampling strategies in Section~\ref{sec:templecontextsampling}. The model pre-trained with TEI gains the best performance, demonstrating the effectiveness of TEI as the pretext task to capture the rich knowledge contained in popularity dynamics for various downstream prediction tasks.

\section{Conclusion}
To the best of our knowledge, we are the first to propose a \emph{pre-training framework for popularity prediction}, which can be effectively transferred into different popularity prediction tasks. We design a novel \emph{temporal elapse inference} as the pretext task for pre-training, impelling the pre-trained model to effectively capture characteristics of popularity dynamics. Experiments conducted on two real datasets with various downstream tasks demonstrate both the effectiveness and generality of the pre-trained model. In the future, we aim to extend the pre-training framework to more scenarios and replace the TCN model with more advanced deep models that also consider user, content, and structure information.

%%
%% The acknowledgments section is defined using the "acks" environment
%% (and NOT an unnumbered section). This ensures the proper
%% identification of the section in the article metadata, and the
%% consistent spelling of the heading.
\begin{acks}
This work is funded by the National Natural Science Foundation of China under Grant Nos. 62102402, U21B2046, and the National Key R\&D Program of China (2020AAA0105200). Huawei Shen is also supported by Beijing Academy of Artificial Intelligence (BAAI).
\end{acks}

%%
%% The next two lines define the bibliography style to be used, and
%% the bibliography file.

\end{document}